# Optimization of Synthesis Parameters and Superconducting Properties of GdFeAsO$_{1-x}$F$_x$

Mohammad Azam[1], Manasa Manasa[1], Tatiana Zajarniuk[2], Svitlana Stelmakh[1], Tomasz Cetner[1], Andrzej Morawski[1], Andrzej Wiśniewski[2], Shiv J. Singh[1*]

[1]Institute of High Pressure Physics (IHPP), Polish Academy of Sciences, Sokołowska 29/37, 01-142 Warsaw, Poland.
[2]Institute of Physics, Polish Academy of Sciences, aleja Lotników 32/46, 02-668 Warsaw, Poland.

*Corresponding Email: sjs@unipress.waw.pl

## Abstract

*RE*FeAsO (*RE*1111; *RE*: rare earth) belongs to the 1111 family of iron-based superconductors (FBS), which illustrates the enhancement of the superconducting transition ($T_c$) with smaller radii of *RE*. However, the synthesis of the 1111 phase with a heavy rare-earth is always challenging. In this paper, we report the optimization of the growth and superconducting properties of F-doped GdFeAsO$_{1-x}$F$_x$ bulks by preparing the samples in a wide temperature range (700-1100°C) at ambient pressure. The optimized synthesis parameters are concluded based on structural, microstructural, transport, and magnetic measurements. These findings suggest that the optimal conditions for preparing F-doped Gd1111 bulks involve a two-step process at 900°C for 61 hours at ambient pressure, which is lower than previously reported. The optimized samples have revealed the superconducting transition temperature ($T_c^{onset}$) of 43 K for GdFeAsO$_{0.83}$F$_{0.17}$. The first-time reported critical current $J_c$ value for this Gd1111 is observed of the order of 10$^3$ A/cm$^2$ at 0 T and 5 K. Our investigation also concluded that highly pure precursors, particularly gadolinium metal, are required to achieve the superconducting properties of F-doped Gd1111. A high growth pressure of 1 GPa reduces the superconducting properties of F-doped Gd1111.

## I. INTRODUCTION

Iron-based superconductor (FBS) was discovered in 2008 through F-doped LaFeAsO with a critical transition temperature $T_c$ of 26 K [1]. Following this pioneering work, more than 100 compounds have been reported that can be categorized into mainly six families based on the structure of the parent compounds: 1111 (*RE*FeAsO), 122 (*A*Fe$_2$As$_2$; *A* = Ba, K)[2], 1144 (*AeA*Fe$_4$As$_4$, *Ae* = Ca; *A* = K), 111 (*AE*FeAs; *AE* = Li, Na), 11 (FeSe), thick perovskite-type oxide blocking layers: 42622 (Sr$_4$V$_2$O$_6$Fe$_2$As$_2$, Sr$_4$Sc$_2$O$_6$Fe$_2$P$_2$) [3]-[5]. 1111 family is always fascinating and provides the highest superconducting transition temperature of 58 K for F-doped Sm1111 [6]. The replacement of *RE* from La to a smaller ionic radius of rare earth iron (*RE*$^{+3}$) generally enhances the superconducting transition [7], [8]. However, it is always challenging to prepare high-quality single-crystal and bulk samples, especially at ambient pressure [9]-[11].

Until now, very few studies have been reported based on the GdFeAsO system, where the parent compound does not depict the superconducting transition but shows the structural and magnetic transition around 150 K [12]. Superconductivity can be induced by various kinds of doping as reported so far, such as fluorine doping at O-sites [12-13], thorium (Th) at Gd-sites [14][15], Ir at Fe sites [16]. In all these dopants, fluorine doping is a very common way to induce the superconductivity with high $T_c$ value in the 1111 family [6]. The transition temperature for F-doped Gd1111 (GdFeAsO$_{1-x}$F$_x$) has been reported around 36 K [13], 45 K [12] and 40 K [17] for 17%, 14% and 20% F-doping, respectively. However, most reported papers have employed a two-step synthesis at high heating temperatures, *i.e.*, ~1200-1300°C [12-13],[17], which is typically unsuitable for the lighter elements like fluorine because of its evaporation problems and it generates many impurity phases, as discussed for 1111 family [9]. Furthermore, this high heating temperature could also result in the explosion of the quartz tube during the sample growth process. These might be reasons for the presence of impurity phases in the reported studies of F-doped GdFeAsO [11],[14],[16-17]. To overcome these problems, we urgently need a robust synthesis process by optimizing the growth process of F-doped GdFeAsO, so that the sample quality and its superconducting properties can be improved and explored. This is our main motivation behind this research paper.

To optimize the superconducting properties of GdFeAs(O,F), a series of bulk samples were prepared by considering various synthesis temperatures and times as well as the effect of Gd metal precursor provided by the different suppliers (ABCR and Alfa Aesar). Structural, microstructure, transport, and magnetic measurements have been performed to conclude the final findings. The optimized synthesis parameters of 900°C and 61 h showed the highest $T_c$ of 43 K and are lower than those of the previous studies [13-17]. Using these optimized growth conditions, a series of F-doped GdFeAsO$_{1-x}$F$_x$ ($x$ = 0.17, 0.2, and 0.3 as a nominal composition) have been prepared and various superconducting properties have been studied. We have also prepared 17% F-doped Gd1111 at 1 GPa by high gas pressure and high-temperature growth (HP-HTS) method [18], suggesting more studies are needed in this direction.

## II. EXPERIMETAL

GdFeAsO$_{1-x}$F$_x$ bulks have been prepared by the starting precursors such as Fe (purity 99.99%), As (purity 99.999%), FeF$_3$ (purity, 99%), and Fe$_2$O$_3$ (purity 99.85%). Since Gd metal is very air-sensitive, we have used Gd metal as a precursor from three suppliers: Gd powder (Alfa Aesar, 99.9%), Gd powder (ABCR, 99.9%), and Gd pieces (Alfa Aesar, 99.9%). To optimize the growth conditions, one composition, GdFeAsO$_{0.83}$F$_{0.17}$, was prepared by using the initial precursors Gd, As, FeF$_3$, and Fe$_2$O$_3$ and heated at different temperatures from 700 to 1100°C for 40 h by sealing in an evacuated quartz tube. In the next step, these samples were opened inside the glove box, reground, pelletized, and then sealed again in an evacuated quartz tube. This prepared tube was heated again at the respective temperature for 21 hours as a two-step process [12-13,17]. One can note that some samples, except those with a synthesis temperature of 900°C, were reground and heated again for longer heating times due to the presence of Gd pieces. After finding the best conditions, various samples with different nominal F-doping levels ($x$) were prepared by this growth process. High-pressure synthesis samples at 1 GPa have been performed at 900°C for 1 h by using HP-HTS as an *ex-situ* process, as discussed elsewhere in more detail [19].

XRD measurements have been performed using an X'Pert PRO, Panalytical diffractometer with filtered Cu–K$\alpha$ radiation (wavelength: 1.5418 Å, power: 30 mA, 40 kV) and a PIXcel$^{1D}$ position scintillation detector. Zeiss Ultra Plus field-emission scanning electron microscope equipped with the EDS microanalysis system by Bruker mod. Quantax 400 with an ultra-fast detector was carried out for the detailed macrostructural analysis and the mapping of the constituent elements. Magnetic and resistivity measurements have been performed by Quantum Design PPMS and the four-probe method using a closed-cycle refrigerator (CCR), respectively.

## III. RESULTS AND DISCUSSION

The first batch of GdFeAsO$_{0.83}$F$_{0.17}$ bulks was prepared by Gd pieces (Alfa Aesar) by using various heating temperatures (700-1100°C) by a two-step process (40 h + 21 h). The structural analysis of these samples depicted the main phase of a tetragonal ZrCuSiAs-type structure[12-13] (Fig. S1), which is not shown here, and their resistivity behavior is revealed in Fig. 1(a). It clearly indicates that the samples prepared at the heating temperature of 900°C depict the superconductivity, whereas other samples prepared at lower or higher heating temperatures than 900°C behave like a parent compound, GdFeAsO with a structure and magnetic transition, as reported elsewhere [12]. The sample prepared at 700°C has this transition at a lower temperature than that reported for GdFeAsO, suggesting a small amount of fluorine inside the lattice. A further increase in the heating temperature, *i.e.*, 800°C, has almost linear resistivity, but the superconductivity is not observed up to 7 K. Similar resistivity behaviors are also obtained for the samples prepared at 1000°C which can suggest a slightly higher amount of the actual fluorine substitution in these samples compared to that of a sample prepared at 700°C, but it's not sufficient to depict the superconducting property [20]. Furthermore, this transition is more clearly visible for the sample prepared at 1100°C, which suggests a reduced amount of fluorine inside the lattice, and its resistivity behavior is similar to that of the parent compound, GdFeAsO. It could be possible due to the evaporation of fluorine from the Gd1111 phase at this high temperature (i.e. 1100°C). In other words, there is not a sufficient amount of fluorine inside the lattice to achieve the superconducting properties for the samples prepared at 700, 800, 1000, and 1100°C. However, the samples prepared at 900°C, 61 h (optimized conditions) have the highest reported value of $T_c^{onset}$~43-44 K and $T_c^{offset}$~26 K. It implies that these conditions are appropriate for F-doping inside the lattice and are lower than the previous reports [11-12][14][16-17][20]. During the growth process, it's very important that small Gd pieces must be covered by other precursors. These studies suggest that the best heating conditions for F-doped Gd1111 are 900°C, (40 h + 21 h) by a two-step process and a very high heating temperature of ~1100-1300°C, as reported earlier [11][12-14][16-17], is not really required.

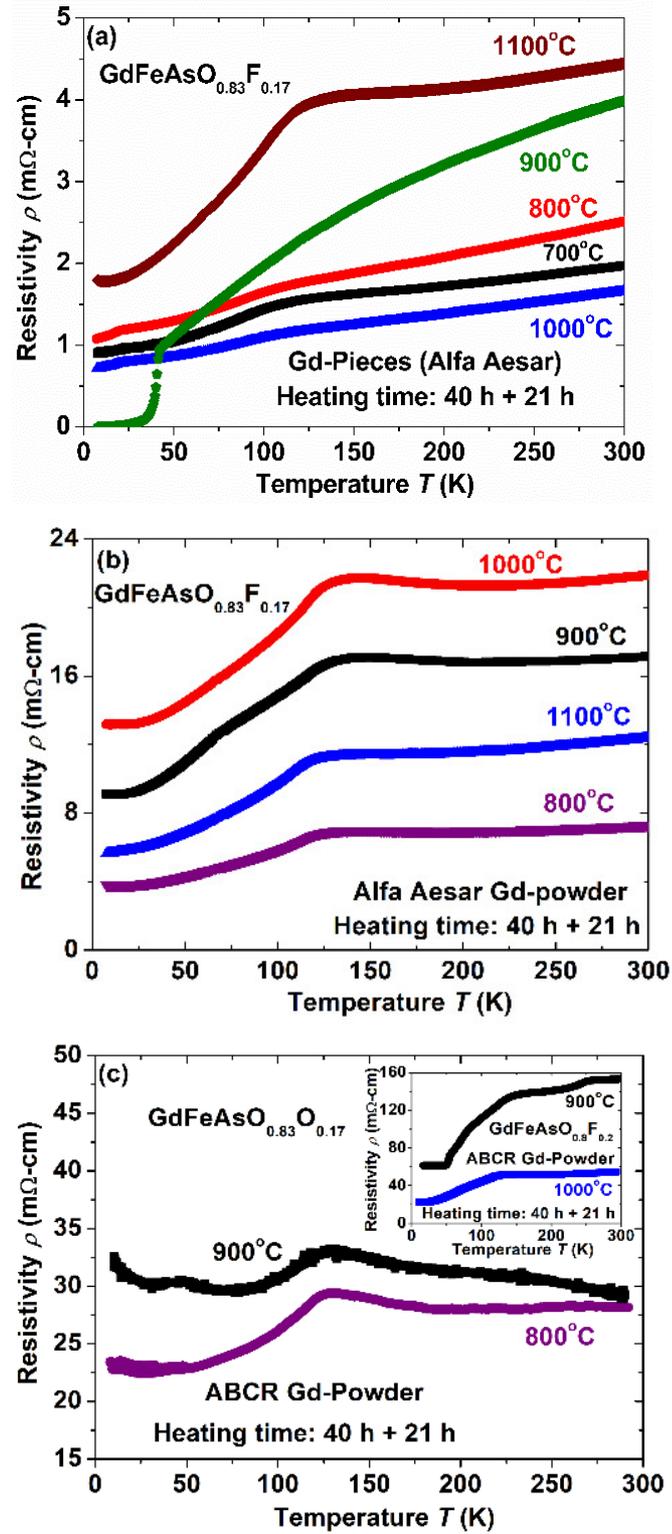

**Fig. 1.** The temperature dependence of resistivity behavior of $GdFeAsO_{0.83}F_{0.17}$ prepared at different temperatures from **(a)** Gd-pieces (Alfa Aesar) **(b)** Gd-powder (Alfa Aesar) **(c)** Gd-powder (ABCR). The inset shows the resistivity behavior for $GdFeAsO_{0.8}F_{0.2}$ prepared by Gd-powder (ABCR).

To reduce the problem with Gd pieces, we used Gd powder (purity 99.9%) from Alfa Aesar in place of Gd pieces. First, GdFeAsO$_{0.83}$F$_{0.17}$ was prepared using Gd powder by following the optimal growth conditions (900°C, 61 h) as discussed above. Unfortunately, the superconductivity was not observed for these samples, as shown in Fig. 1(b). To be more specific, GdFeAsO$_{0.83}$F$_{0.17}$ bulks were also prepared using this Gd powder (Alfa Aesar) at various temperatures, including 1000°C, 1100°C, and 800°C, but still, the superconductivity was not observed as depicted in Fig. 1(b). These samples have a similar resistivity behavior as that of the parent compound [11-12] and a significant amount of the GdOF impurity phase, suggesting that there is not a sufficient amount of F-content inside the lattice to induce superconductivity [11-12]. To reach the final conclusions, we prepared the increased fluorine-doped contents, such as 20%, 30%, and 40%, at 900°C for 61 hours(supplementary Fig. S1 and S2). The resistivity measurements of these samples are depicted in Fig. S2, where the superconductivity was not observed. All these samples behave very similarly to the parent compound, but the explosion occurred more frequently inside the furnace. The properties of all these samples suggest that Gd powder from Alfa Aesar could have some impurities. Since 1111 phase formation is very sensitive to the purity of precursors, especially at the first step, as reported in various studies [6][9][11]. It could be a reason that we have not observed a superconductivity for GdFeAsO$_{0.83}$F$_{0.17}$ bulks using Alfa Aesar powder.

In the next step, Gd powder is employed from a different supplier, *i.e.*, ABCR Germany, which can provide the highest purity of 99.9%. For a quick check of the quality of this powder, two GdFeAsO$_{0.8}$F$_{0.2}$ and GdFeAsO$_{0.83}$F$_{0.17}$ samples were prepared at 900°C, 800°C, and 1000°C for 61 hours, by a two-step process as mentioned above. The temperature dependence of the resistivity behaviors of these samples is shown in Fig. 1(c). Frequently, an explosion inside the furnace occurred during the growth process. However, the superconductivity was not observed as cleared from Fig. 1(c), and all samples behave like a parent compound. These results suggest that Gd powder from ABCR is also not suitable for the growth of F-doped Gd1111. Hence, we can clearly conclude that Gd powder, even from different suppliers, as a precursor for the superconducting phase formation of F-doped Gd1111 is not appropriate, even though it has the same purity as that of Gd pieces.

With these optimized growth conditions and using Gd pieces, a series of F-doped GdFeAsO$_{1-x}$F$_x$ have been prepared, and their XRD patterns are depicted in Fig. 2. All samples have the main tetragonal ZrCuSiAs-type structure with the lattice parameters of $a$ = 3.909(2) Å and $c$ = 8.441(3) Å for GdFeAsO$_{0.83}$F$_{0.17}$. Further, by increasing the concentration of fluorine (*x*), the estimated lattice parameters are to be almost the same as in the previous studies [11],[13] where samples were prepared at very high heating temperatures. One can note that due to the impurity phases, there can be a large error in the calculation of the lattice parameters. In addition to the main 1111 phase, a small amount of GdAs and GdOF were also observed, which were enhanced with higher concentrations of fluorine contents, *i.e.*, for *x* = 0.2 and 0.3. These impurity phases are similar to the previous studies based on F-doped Gd1111 [11-13]. Our studies confirm that the low heating conditions (900°C, 61 h) by a two-step synthesis are sufficient for the phase formation of F-doped Gd1111, however, the initial precursors especially Gd metal, must be highly pure to observe a superconducting Gd1111 phase with F-doping.

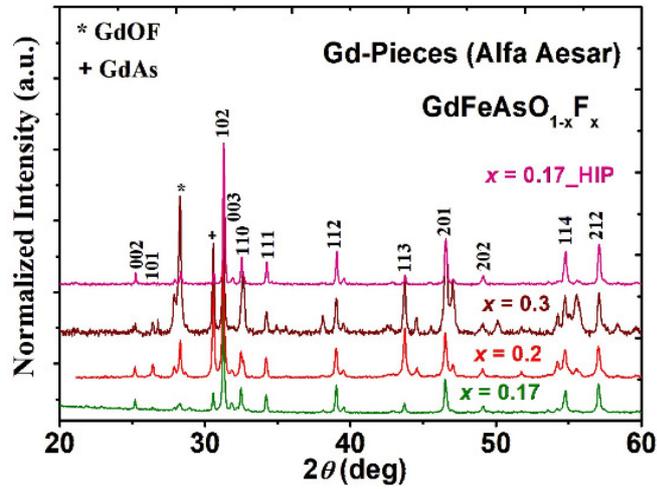

**Fig. 2.** The XRD patterns of various F-doped GdFeAsO$_{1-x}$F$_x$ prepared by the optimized conditions (900°C, 40 h + 21 h).

Mapping and backscattered electrons (BSE) images for Gd1111 for *x* = 0.17, 0.3, and 0.17_HIP are shown in Fig. 3. Black, white, and gray contrasts are observed for GdAs or pores, Gd$_2$O$_3$, and Gd1111 phases. These images suggest that impurity phases are enhanced with increased fluorine doping at ambient pressure and also through HP-HTS. The depicted mapping for *x* = 0.3, 0.17_HIP also supports the enhancement of inhomogeneity of the constituent elements compared to 17% F-doped samples (*x* = 0.17). It confirms that the sample *x* = 0.17 has an almost homogeneous distribution of constituent elements, whereas higher fluorine doping enhances the inhomogeneity of the constituent elements, suggesting the presence of GdOF and GdAs phase (more details mapping as a supplementary Fig. S3). It supports the above-discussed XRD data analysis.

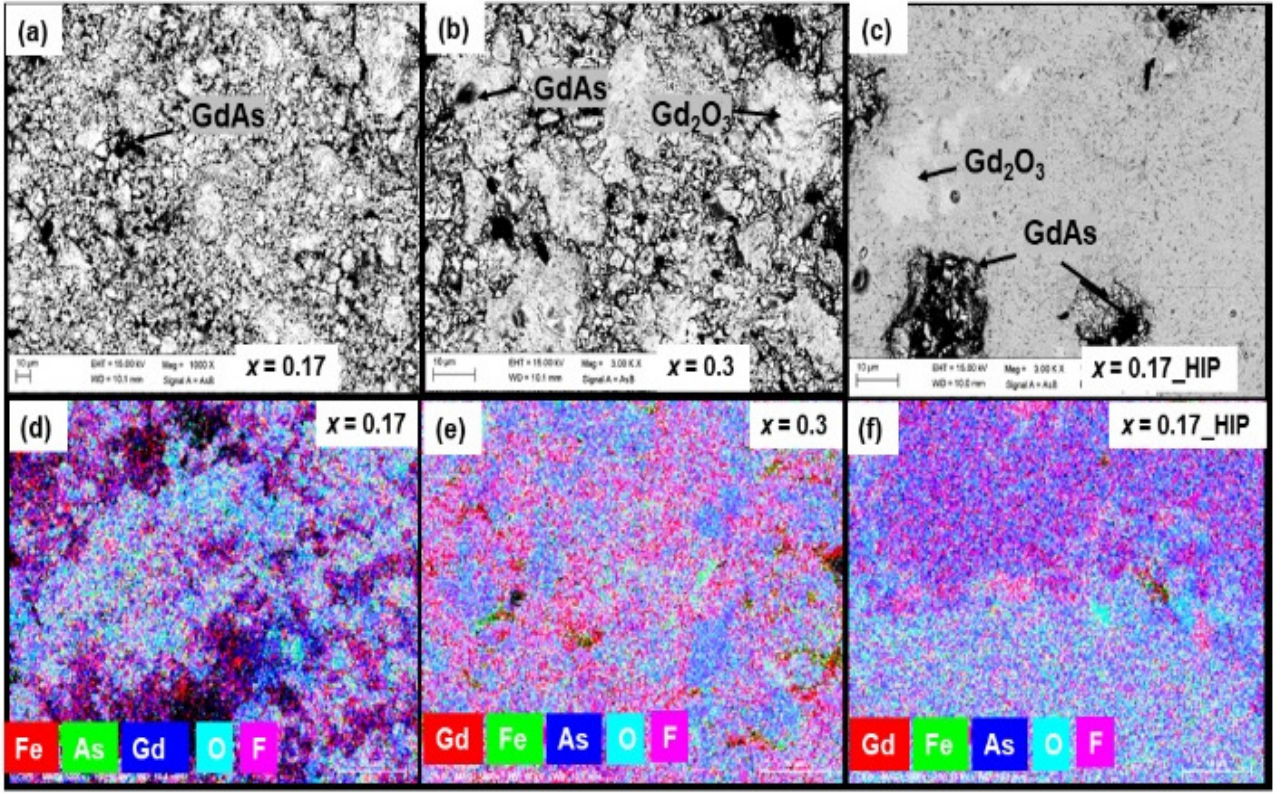

**Fig. 3.** Back-scattered image and elemental mapping of all constituent elements for GdFeAsO$_{1-x}$F$_x$: **(a),(d)** for $x = 0.17$; **(b),(e)** for $x = 0.3$ and **(c), (f)** for $x = 0.17$_HIP prepared at 1 GPa.

The temperature dependence of the resistivity behavior for these samples ($x = 0.17$, 0.2, 0.3, and 0.17_HIP) is depicted in Fig. 4. The normal state behavior of all these samples is metallic, whereas the resistivity value is slightly higher for the samples with a large number of impurity phases, as presented in Fig. 4(a). One can observe the superconducting transition for all samples. To present more clearly, the low-temperature resistivity behavior of these samples is revealed in Fig. 4(b). The sample $x = 0.17$ shows the highest transition temperature ($T_c^{onset}$) of 43 K, and a wider transition (~18 K) is observed, which is similar to the previous reports [11-13]. Interestingly, the samples $x = 0.2$ and 0.3 depict a slightly lower $T_c^{onset}$ value of around 40 K with a very broad transition width (>22 K), which suggests the enhancement of impurity phases, as discussed above. The $T_c^{onset}$ value of 43-44 K is nearly identical to the highest reported value (~45 K) for F-doped Gd1111 [12]. It implies that, in comparison to the high synthesis temperature, our low-temperature synthesis method reduces the explosion of quartz tubes during the growth process, improving the activity of fluoride and simplifying a reaction between fluorine agents and other precursors[13-16]. The sample $x =$

0.17_HIP has a reduced $T_c^{onset}$, suggesting that the formation of the superconducting phase is not supported by an extremely high pressure (1 GPa), which is in contrast to the previous report based on *in-situ* growth process of F-doped Gd1111 by high-pressure (1350°C, 5 GPa)[11].

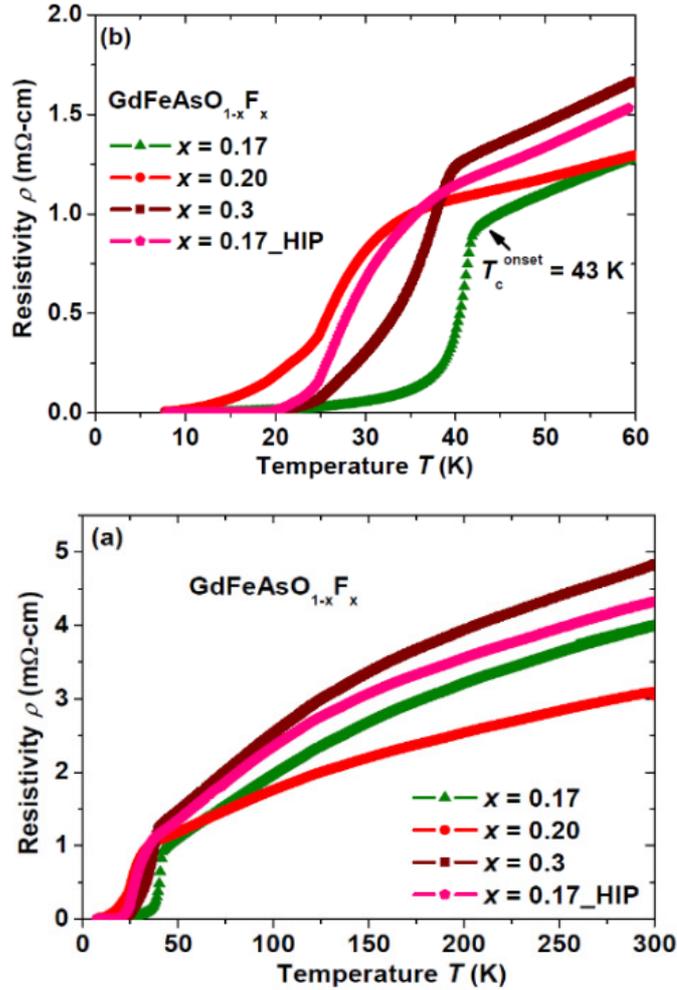

**Fig. 4. (a)** The temperature dependence of the resistivity up to room temperature and **(b)** Low-temperature variation of resistivity up to 60 K for GdFeAsO$_{1-x}$F$_x$.

To confirm the Meissner effect of these samples, Fig. 5(a) represents the temperature dependence of the normalized magnetic moment for these samples. A paramagnetic background was observed due to the presence of the impurity phases [13], as reported for other 1111 members [9]. These graphs depict a magnetic $T_c^{onset}$ value of 42 K, 38 K, 36 K, and ~28 K for the sample $x$ = 0.17, 0.2, 0.3, and 0.17_HIP, respectively. The magnetic transition behavior can be considered as a combined effect of the superconducting phase and the existence of secondary phases, similar to the resistivity measurements.

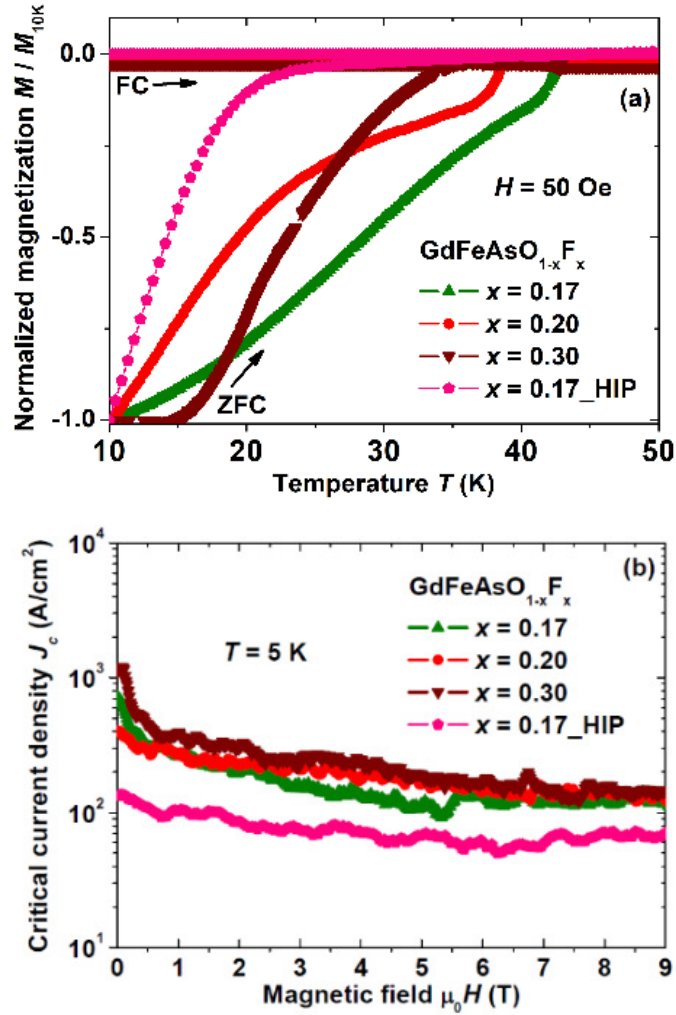

**Fig. 5. (a)** The temperature dependence of the normalized magnetization at 50 Oe and **(b)** The magnetic field dependence of critical current density at $T = 5$ K for GdFeAsO$_{1-x}$F$_x$.

The critical current density ($J_c$) is an important parameter for a superconducting material. Interestingly, there is no report for the $J_c$ property of F-doped Gd1111. We have measured the magnetic hysteresis loop (*M-H*) for these samples, which has been influenced by the paramagnetic phase. The shape of this *M-H* loop is similar to the one reported for FBS materials [5],[6]. To calculate the $J_c$ value, Bean model is applied: $J_c = 20\Delta m / Va(1-a/3b)$ [21], where $\Delta m$ is the hysteresis loop width, $V$ is the volume of the sample, shorter and longer edges of the sample are $a$ and $b$, respectively. The calculated $J_c$ with the magnetic field variation is shown in Fig. 5(b) for $x =$ 0.17, 0.2, 0.3, and 0.17_HIP. Interestingly, all samples have depicted similar behavior, and the highest $J_c$ value of the order of $10^3$ A/cm$^2$ is observed. However, the sample $x = 0.17$_HIP has a lower $J_c$ value than other samples, which could be due to the reduced pinning centers or the presence of impurity phases in between grains or inside grains. One can note that the $J_c$ behavior of Gd1111 depicts almost independence of the magnetic field up to 9 T, indicating a good sign for the practical applications. However, this value is lower than the practical $J_c$ value (>$10^4$ A/cm$^2$)[21]. Hence, more research is required in this direction to optimize the growth conditions by applying advanced techniques for the materials process [9], such as high-pressure synthesis and annealing.

## IV. CONCLUSIONS

The physical properties optimization of F-doped Gd1111 has been performed by considering synthesis temperature and time, as well as the purity of Gd metal from different suppliers. Our study concludes that a two-step process with the optimized conditions of 900°C and 61 hours is sufficient to obtain the high superconducting properties of F-doped Gd1111 with a high $T_c$ value of ~43-44 K at the ambient pressure. The critical current density is observed of the order of $10^3$ A/cm$^2$ at 5 K and 0 T. The properties of the first F-doped Gd1111 prepared by high-pressure synthesis at 1 GPa suggest that very high growth pressure is not supportive for the superconducting properties of F-doped Gd1111. We believe this study will enhance the interest of the research community in further studies of Gd1111 and the formation of superconducting wires and tapes.

## V. ACKNOWLEDGMENTS

This research was funded by National Science Centre (NCN), Poland, grant number "2021/42/E/ST5/00262" (SONATA-BIS 11). S.J.S. acknowledges financial support from National Science Centre (NCN), Poland through research Project number: 2021/42/E/ST5/00262.

Supplementary Data

# Optimization of Synthesis Parameters and Superconducting Properties of GdFeAsO$_{1-x}$F$_x$

Mohammad Azam[1], Manasa Manasa[1], Tatiana Zajarniuk[2], Svitlana Stelmakh[1], Tomasz Cetner[1], Andrzej Morawski[1], Andrzej Wiśniewski[2], Shiv J. Singh[1*]

[1]Institute of High Pressure Physics (IHPP), Polish Academy of Sciences, Sokołowska 29/37, 01-142 Warsaw, Poland.

[2]Institute of Physics, Polish Academy of Sciences, aleja Lotników 32/46, 02-668 Warsaw, Poland.

*Corresponding Email: sjs@unipress.waw.pl

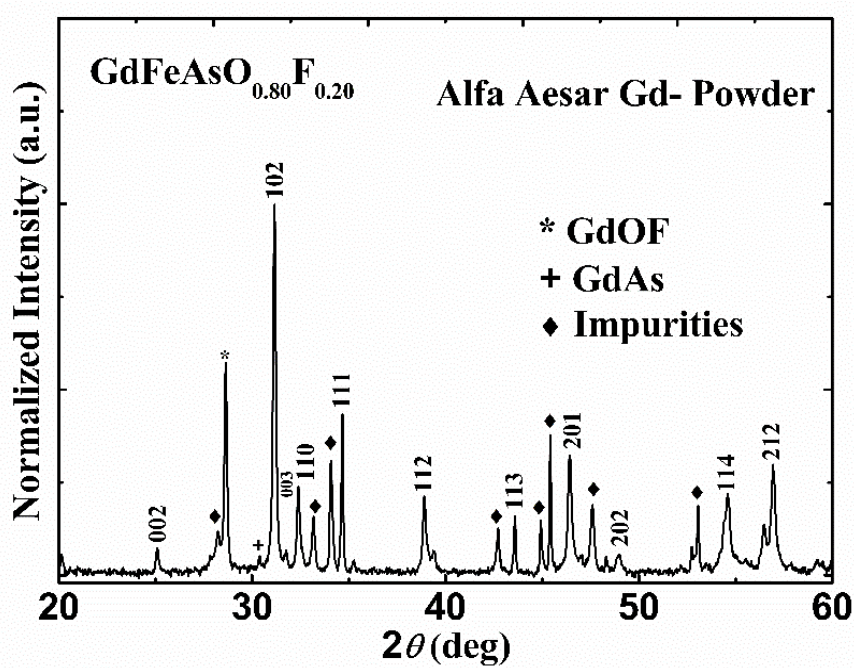

**Fig. S1.** The XRD patterns of GdFeAsO$_{0.80}$F$_{0.20}$ prepared by a two-step process with the conditions (900°C, 40 h + 21 h) using Gd-powder (Alfa Aesar) as a precursor.

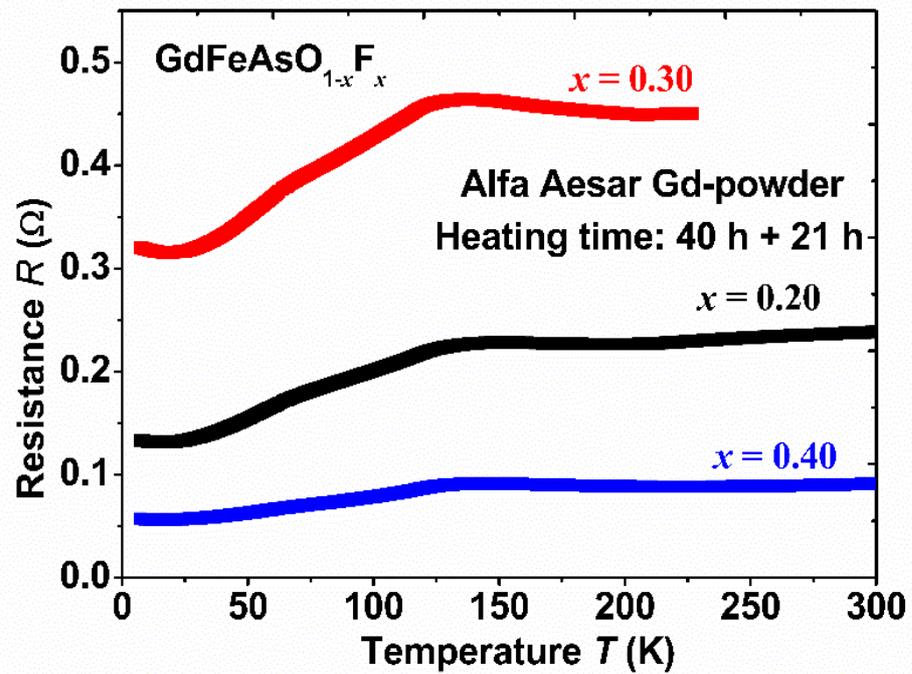

**Fig. S2.** The temperature dependence of resistivity behavior of GdFeAsO$_{1-x}$F$_x$ ( $x$ = 0.20, 0.30, 0.40 as a nominal composition ) prepared by a two-step process with the conditions (900°C, 40 h + 21 h) using Gd-powder (Alfa Aesar) as a precursor.

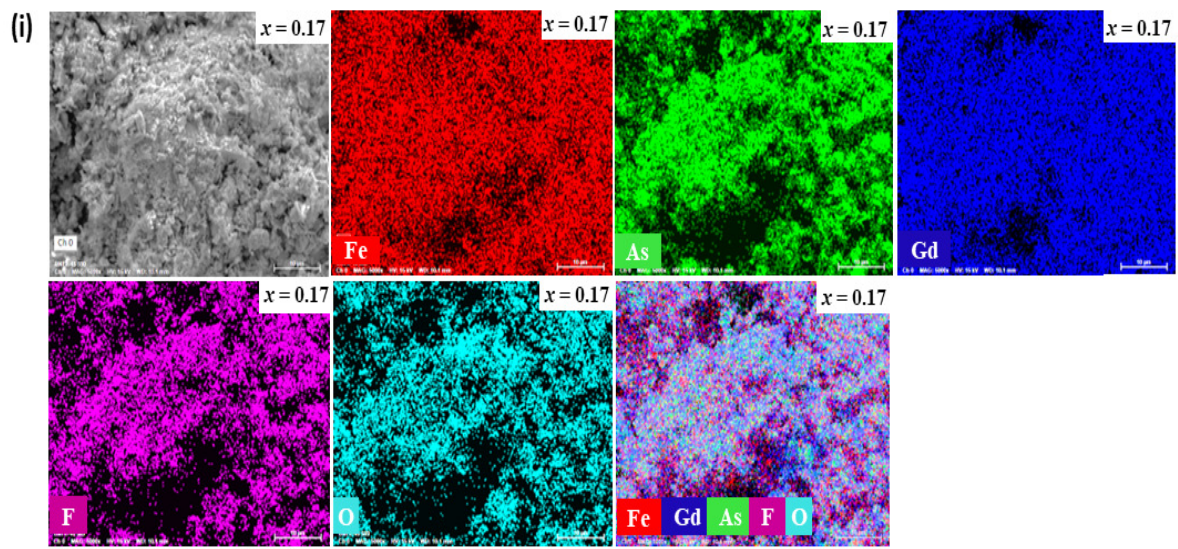

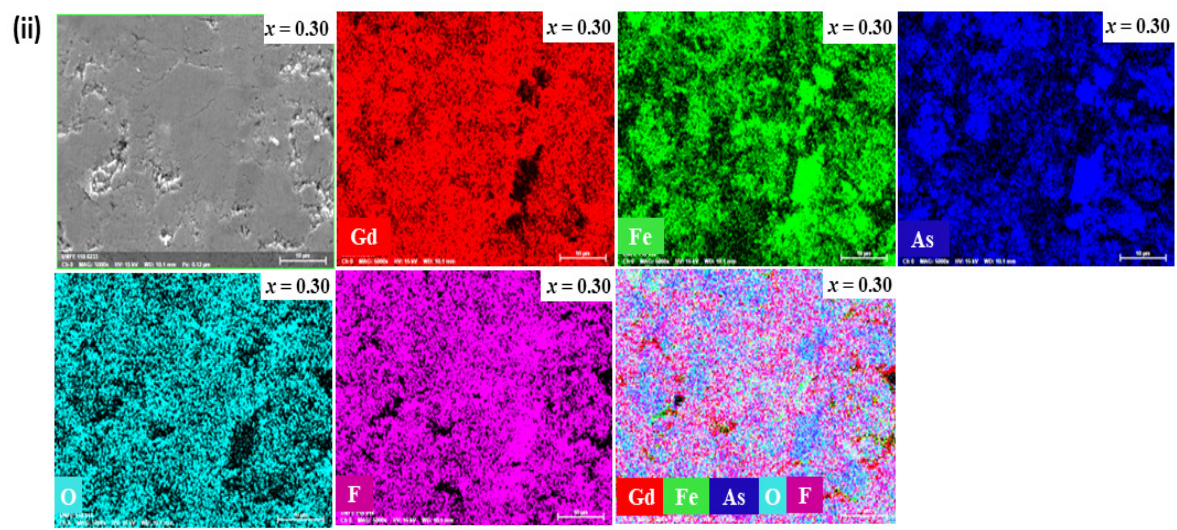

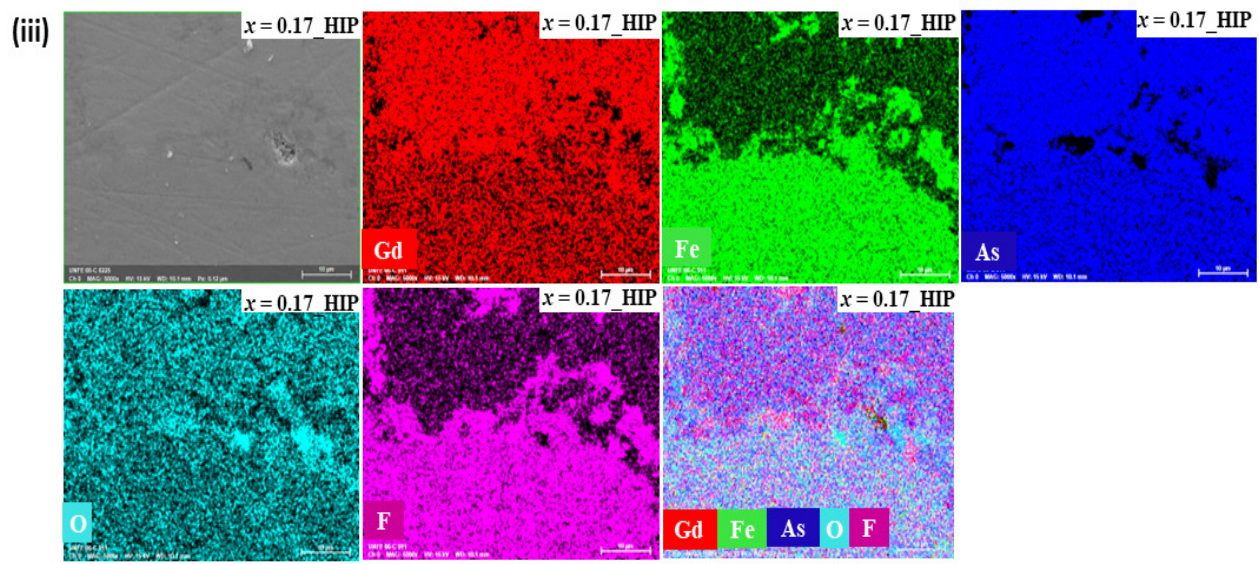

**Fig. S3.** Mapping of all constituent elements of GdFeAsO$_{1-x}$F$_x$ bulks synthesized by using Gd-pieces (Alfa Aesar) as a precursor for: **(i)** $x$ = 0.17 **(ii)** $x$ = 0.3 prepared by conventional synthesis process at ambient pressure (CSP); and **(iii)** $x$ = 0.17_HIP prepared at 1 GPa.